\documentclass[showpacs,twocolumn,pra]{revtex4-1}

\usepackage{slashed}
\usepackage{mathrsfs}
\usepackage{amsfonts}
\usepackage{amsmath}
\usepackage{amssymb}
\usepackage{revsymb}
\usepackage{graphicx}
\usepackage{mathrsfs}
\usepackage{bm}
\usepackage{psfrag}
\usepackage{color}
\usepackage{hyperref}
\usepackage{natbib}
\usepackage{bbm}
\usepackage{bbold}

\usepackage[usenames,dvipsnames]{xcolor}

\newcommand{\be}{\begin{equation}}
\newcommand{\ee}{\end{equation}}
\newcommand{\ba}{\begin{array}}
\newcommand{\ea}{\end{array}}
\newcommand{\bea}{\begin{eqnarray}} 
\newcommand{\eea}{\end{eqnarray}} 
\newcommand{\bd}{\begin{displaymath}}
\newcommand{\ed}{\end{displaymath}}
\newcommand{\eps}{\varepsilon}

\newcommand{\trm}[1]{\textrm{#1}}

\newcommand{\av}[1]{\langle #1 \rangle}
\newcommand{\figref}[1]{Fig. \ref{#1}}

\newcommand{\eqnref}[1]{Eq. (\ref{#1})}

\newcommand{\vphi}{\varphi}
\newcommand{\vphid}{\varphi_{\Delta}}
\newcommand{\vphis}{\varphi_{\Sigma}}
\newcommand{\xis}{\xi_{\Sigma}}
\newcommand{\ks}{k_{\Sigma}}
\newcommand{\kd}{k_{\Delta}}

\newcommand{\kbar}{\bar{k}}
\newcommand{\vphib}{\bar{\vphi}}

\newcommand{\checked}{}
\newcommand{\checkedb}{}
\newcommand{\am}{\trm{am}}

\newcommand{\phpol}{\textrm{e}}
\newcommand{\pol}{\eps}
\newcommand{\epol}{\trm{e}}

\newcommand{\nn}{\nonumber}
\newcommand{\pin}{p_{\tiny\trm{in}}}
\newcommand{\tin}{\tiny\trm{in}}
\newcommand{\Pin}{\Pi_{\trm{in}}}
\newcommand{\e}{\mathbb{e}}

\newcommand*\xbar[1]{%
  \hbox{%
    \vbox{%
      \hrule height 0.5pt 
      \kern0.2ex
      \hbox{%
        \kern-0.15em
        \ensuremath{#1}%
        \kern-0.15em
      }%
    }%
  }%
}

\newcommand{\cole}[1]{#1}
\newcommand{\colb}[1]{#1}
\newcommand{\colbk}[1]{#1}
\newcommand{\colbkq}[1]{#1}

\newcommand{\optional}[1]{}

\newcommand{\colH}[1]{#1}
\newcommand{\wegb}[1]{}

\newcommand{\Ben}[1]{}

\begin{document}

\title{Classical and quantum dynamics of a charged scalar particle in a background of two counterpropagating plane waves}
 
\author{B.~King}
\affiliation{Centre for Mathematical Sciences, Plymouth University, Plymouth, PL4 8AA, United 
Kingdom}
\email{b.king@plymouth.ac.uk}

\author{H.~Hu}
\affiliation{Hypervelocity Aerodynamics Institute, China Aerodynamics Research and 
Development Center, 621000 Mianyang, Sichuan, China}


\date{\today}
\begin{abstract}

We consider a scalar particle in a background 
formed by two counter-propagating plane waves. Two cases are studied: i) dynamics at a magnetic node 
and ii) zero initial transverse canonical momentum. The Lorentz and Klein-Gordon equations are 
solved for these cases and approximations analysed. For the magnetic node solution 
(homogeneous, time-dependent electric field), the modified Volkov wavefunction which arises from a 
high-energy approximation is found to be inaccurate for all energies and the solution itself 
unstable when photon emission (nonlinear Compton scattering) is included. For the zero 
initial transverse canonical momentum case, in both quantum and classical cases, forbidden 
parameter regimes, absent in the plane wave model, are identified.
\end{abstract}
\maketitle
%
%
%
\colH{For quantum electrodynamical (QED) calculations in a strong laser background, a general method 
to deal with the \wegb{nonperturbative} interaction between \wegb{the} laser fields and \wegb{the} 
\colb{charged} particles is to employ the laser-dressed particle-state \wegb{which is the} solution 
of the relevant relativistic quantum dynamical equation (Dirac for fermions  and Klein-Gordon (KG) 
for scalars). However, only for \colb{a very limited  number of background fields has} the exact 
solution \colb{been} obtained analytically. The most widely used ``Volkov states'' are the 
solutions to the Dirac and KG equations in a plane-wave electromagnetic background} (reviews can be 
found in \cite{ritus85,marklund_review06,dipiazza12,king15a}). \colb{These form the basis of the 
\emph{plane wave model}. In this model, QED processes with highly relativistic incoming particles in 
an arbitrary laser field background are well-approximated by calculating the same processes in a 
plane-wave background. This is supposed valid when the electromagnetic invariants are much smaller 
than the classical and quantum nonlinearity parameters \cite{narozhny15}.}
\newline

Due to the high degree of spatial focussing required to reach extreme field intensities in experiment, there has been recent interest in going beyond the plane wave model. Univariate, transverse but non-lightlike backgrounds have been studied for the case of $k^{2}>0$ (an electric vacuum) \cite{becker77,mendonca11,raicher13,varro13,varro14,raicher15} and $k^{2}<0$ (a magnetic vacuum) \cite{cronstroem77,becker77,king16b}. Motivation for calculating QED in non-lightlike backgrounds stems from interest in quantum processes in dispersive media such as crystals \cite{uggerhoj5} and plasmas \cite{dipiazza07} but also strong magnetic backgrounds such as found in astrophysical objects like magnetars \cite{harding06}.
\newline

The constructive interference that accompanies coherent addition of multiple laser pulses has often been suggested as a mechanism to reach the high field intensities required to trigger electron-positron cascades in an experiment \cite{bulanov10b,gelfer15}. On the one hand, the magnetic node of a standing wave is a particularly popular background for simulations \cite{kirk08,nerush11,elkina11,king13a,mironov14}, which rely upon the locally constant field approximation \cite{king13b,harvey15} within the plane wave model. On the other hand, there is a rich particle dynamics even when just two plane waves are combined to form a standing wave (this has recently been investigated classically when radiation reaction is incorporated \cite{kirk16}). 

High-energy approximations for the wavefunction of scalar charged particles in a standing-wave  
background have recently been acquired \cite{hu15}, and corresponding deviations from the plane 
wave model for scalar nonlinear Compton scattering in a standing wave \cite{raicher16} \cole{and 
nonlinear Breit-Wheeler pair production in a focussed beam \cite{dipiazza16} have been suggested}.
\newline

In the current paper, we solve the Lorentz and Klein-Gordon equations analytically to obtain the classical and quantum dynamics for a charged scalar particle in the background of two counter-propagating plane waves. Two solutions are presented: i) particle dynamics at a magnetic node and ii) dynamics for a particle with zero initial transverse canonical momentum.
\newline

The paper is organised as follows: in Sec. I the solution to the Lorentz equation is presented and some example particle trajectories plotted; in Sec. II solutions to the Klein-Gordon equation are presented; in Sec. III high-energy and WKB approximations to the KG equation are discussed and in Sec. IV the approximations are evaluated by comparing their quasi-momentum to the exact solution. The paper is then concluded in Sec. V. \colH{Natural units $\hbar = c = 1$ are employed throughout the paper.}
%
%
%
\section{Classical Dynamics}
\colH{The Lorentz equation for a charge $e>0$ with four-momentum $p$ and mass $m$ in an electromagnetic (EM) field with field tensor $F$ is}
\[
 \dot{p}^{\mu}  = \frac{e}{m} F^{\mu\nu} p_{\nu},
\]
where a dot represents differentiation with respect to the proper time $\tau$. Let $a=eA$ be the scaled vector potential, written as the sum of two plane waves: $a = a_{1}(\vphi_{1}) + 
a_{2}(\vphi_{2})$, where $\vphi_{j} = k_{j} \cdot x$, $j\in\{1,2\}$ with wavevectors 
satisfying $k_{j}\cdot k_{j}=k_{j}\cdot a_{j}=0$. Then the Lorentz equation can be written as
\[
 \dot{\Pi}^{\mu} = \frac{p\cdot \dot{a}_{1}}{p\cdot k_{1}}~k_{1}^{\mu} + \frac{p\cdot 
\dot{a}_{2}}{p\cdot k_{2}}~k_{2}^{\mu},
\]
where $\Pi = p+a$ is the \emph{canonical momentum}. Forming the scalar product of both sides of the equation with $p_{\mu}$, it is clear that $p\cdot 
p$ is an invariant, as expected for a particle on the mass shell. For an arbitrary constant 
four-vector $\pol^{\mu}$, we find another conservation law:
\bea
\frac{d}{d\tau}\left[\left(\pol - \frac{\pol\cdot k_{2}}{k_{1}\cdot k_{2}}~k_{1}- 
\frac{\pol\cdot k_{1}}{k_{1}\cdot k_{2}}~k_{2}\right)\cdot \Pi\right] =0. \label{eqn:class0}
\eea\vspace{0.1cm}
Combining two equations that are formed when $k_{1}$ and $k_{2}$ are dotted into the Lorentz equation, a final 
``longitudinal'' conservation law can be acquired:
\bea
 \frac{d}{d\tau}\left[2 k_{1}\cdot \Pi~k_{2}\cdot \Pi - k_{1}\cdot k_{2}\,\Pi^{2}\right] = 0. 
\label{eqn:class1}
\eea
One can define a useful four-vector:
\[
 \epol_{l,j} = \pol_{l,j} - \frac{\pol_{l,j}\cdot k_{2}}{k_{1}\cdot k_{2}}~k_{1}- 
\frac{\pol_{l,j}\cdot k_{1}}{k_{1}\cdot k_{2}}~k_{2},
\]
for $l \in\{1,2\}$, where $\pol_{l,j}$ is the $j$th polarisation vector of $a_{l}$. Since $\epol_{l,j}\cdot \Pi$ is conserved and $\epol_{l,j}\cdot k_{1} = \epol_{l,j}\cdot k_{2} = 0$, the 
set $\{\epol_{l,1}, \epol_{l,2}, k_{1}, k_{2}\}$ forms a useful basis.
\newline

One major difficulty in solving these sets of equations is encountered when seeking a separable solution. Consider the case of two counter-propagating
circularly-polarised plane waves of the form:
\bea
a_{l} &=& m\xi_{l}\left[\eps_{l,1} \cos\vphi_{l} + \eps_{l,2} 
\sin{\vphi_{l}}\right], \label{eqn:a1}
\eea
where again $l \in\{1,2\}$. If the longitudinal equation, \eqnref{eqn:class1}, is solved first, the external field phases' dependency on the proper time can be used to solve the remaining equations.
Defining the combinations $\vphi_{\Delta} = \vphi_{1}-\vphi_{2}$, $\vphi_{\Sigma} = 
\vphi_{1}+\vphi_{2}$, for counter-propagating waves we see that terms quadratic in the potential can be written $-a\cdot a/m^{2} =\xi_{\Sigma}^{2} - 4\xi_{1}\xi_{2}\sin^{2}(\vphi_{\Delta}/2)$, which is just a function of $\vphi_{\Delta}$, whereas terms linear in the potential can only be written as products of functions of both phase variables (subscripts $\Sigma$ ($\Delta$) correspond to adding (subtracting) the quantity from \Ben{this ``from'' has to be here} $a_{2}$ to (from) the quantity from $a_{1}$.) \Ben{I've not changed this because the definition of $\xis$ occurs in the same sentence. It think this is fairly obvious for the reader - any ``$\Delta$'' or ``$\Sigma$'' symbols are thus defined by this sentence.}
 Elimination of either the quadratic or the linear term greatly 
simplifies analysis, and even the particle dynamics in two non-counter-propagating plane waves can be solved in this case.
\newline

To best demonstrate the main issues involved, let us now specialise to a head-on collision of plane 
waves, meaning $k_{1}\cdot a_{2} = k_{2}\cdot a_{1} = 0$ and $\eps_{l,j} = \eps_{j}$.  Then 
\bea
 \dot{\vphi}_{\Delta}^{2} - \dot{\vphi}_{\Sigma}^{2} = \frac{(\kd\cdot 
\Pin)^{2}\!}{m^{2}}-\frac{(\ks\cdot \Pin)^{2}\!}{m^{2}}
+ 2k_{1}\cdot k_{2}\! \frac{\Pi^{2}-\Pi_{\trm{in}}^{2}}{m^{2}} \nn \label{eqn:univ1}\\
\eea
where quantities with subscript ``$\trm{in}$'' correspond to initial values and $\Pi = 
\Pi(\vphid,\vphis)$. The field-dependent part of the canonical momentum reduces to the Volkov 
exponent in the plane-wave limit of 
$k_{2}\to k_{1}$, $a_{2} \to 0$. For general initial conditions, the canonical momentum term is not 
separable. 
However, we can define two cases for which \eqnref{eqn:class1} yields an analytical solution. 
\newline

\emph{Transverse motion at a magnetic node} can be acquired by choosing the initial electron 
momentum to be entirely transverse and setting $\vphi_{\Delta} = 0$, implying $\vphi_{1} = 
\vphi_{2} = \vphi$ and $\Pi=\Pi(\vphi)$. This reduces \eqnref{eqn:univ1} to a univariate ODE. To 
answer the question of whether this solution is stable, let us pick a frame in which 
$\omega_{1}=\omega_{2}=\omega$ and choose $z$ as the propagation axis. Then if the plane waves are 
counter-propagating, the condition $\vphi_{\Delta}=0$ implies $z=0$, which we use to define the 
position of the magnetic node (other nodes are available at $\omega z = n\pi$ for 
$n\in\mathbb{Z}$). The particle will remain at this longitudinal position if the additional 
condition $\xi_{1}=\xi_{2}$ is fulfilled. Then it can be shown that $\ddot{\vphi}_{\Delta}\propto 
\sin(\vphi_{\Delta}/2)$ and $\dot{\vphi}_{\Delta}\propto p_{z}$. Under these conditions, if 
$p_{z}=0$ when $z=0$, the particle will remain at $z=0$ for all time. This implies the phase can be 
written $\vphi = \omega t = \kbar \cdot x :=\vphib$, where $\kbar = \omega(1,0,0,0)$ is a timelike 
wavevector. We then note $-a^{2} = m^{2}\xis^{2}$ and $p\cdot a$ can be rewritten using \eqnref{eqn:class0} as:
\bea
2 p \cdot a + a^{2} = 2 \Pi_{\tin} \cdot a - a^{2}. \label{eqn:ptransid}
\eea
to give:
\bea
 \left(\frac{d\widetilde{\vphi}}{d\tau}\right)^{2} = \varpi_{\perp}^{2} 
 -\mu_{\perp}^{2}
\sin^{2}\!\frac{\widetilde{\vphi}}{2}\, \label{eqn:perptrajeqn1} \checked
\eea
\[
\varpi_{\perp}^{2} = \frac{(\kbar\cdot 
\Pi_{\tin})^{2}}{m^{2}}+\kbar^{2}\left[\xis^{2}-\frac{\Pi_{\tin}^{2}-m^{2}}{m^{2}}+2\xis\,\frac{
|\Pi_ { \tin } ^ { \perp}|}{m}\right]; \checked
\]
\[
\mu_{\perp}^{2} = \frac{2\kbar^{2}\xis |\Pi_{\tin}^{\perp}|}{m}, \checked
\]
where  $|\Pi^{\perp}_{\tin}|^{2} = (\Pi_{\tin}\cdot \eps_{1})^{2}+(\Pi_{\tin}\cdot \eps_{2})^{2}$, 
 $\widetilde{\vphi} = \vphib-\vphi_{0}$ and $\tan\vphi_{0} = 
\Pi_{\tin}\cdot \eps_{2}/\Pi_{\tin}\cdot \eps_{1}$. 
\eqnref{eqn:univ1} can be directly integrated to give:
\[
 \vphib = 2\,\am\left( 
\frac{\varpi_{\perp}\tau}{2} 
\Bigg| ~\frac{\mu_{\perp}^{2}}{\varpi_{\perp}^{2}}  \right) + \vphi_{0},
\]
where $\am(\cdot|\cdot)$ is the Jacobi amplitude function \cite{nist_dlmf}. 
A further use of \eqnref{eqn:class0} yields an analytical solution to the transverse co-ordinates, but since this adds little to the discussion, it has not been included. Four conserved momenta were identified in this type of background, but since motion at a magnetic node is confined to a plane, it sufficies to use just the transverse degrees of freedom \eqnref{eqn:class0} and the on-shell condition $p^{2}=m^{2}$.
\begin{figure}[!h] 
\centering
 \includegraphics[draft=false,width=4.3cm]{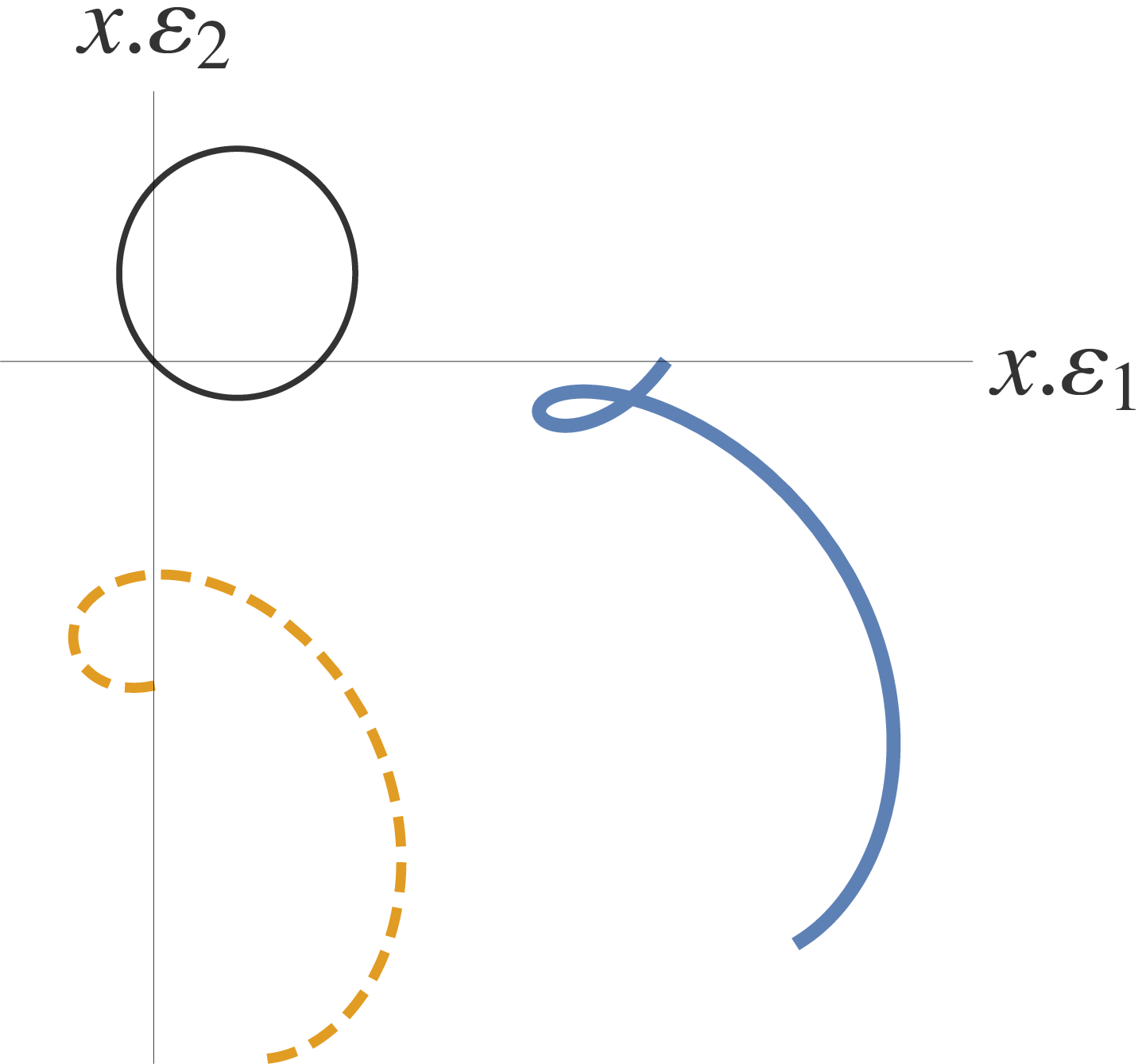}\hfill
 \includegraphics[draft=false,width=3.7cm]{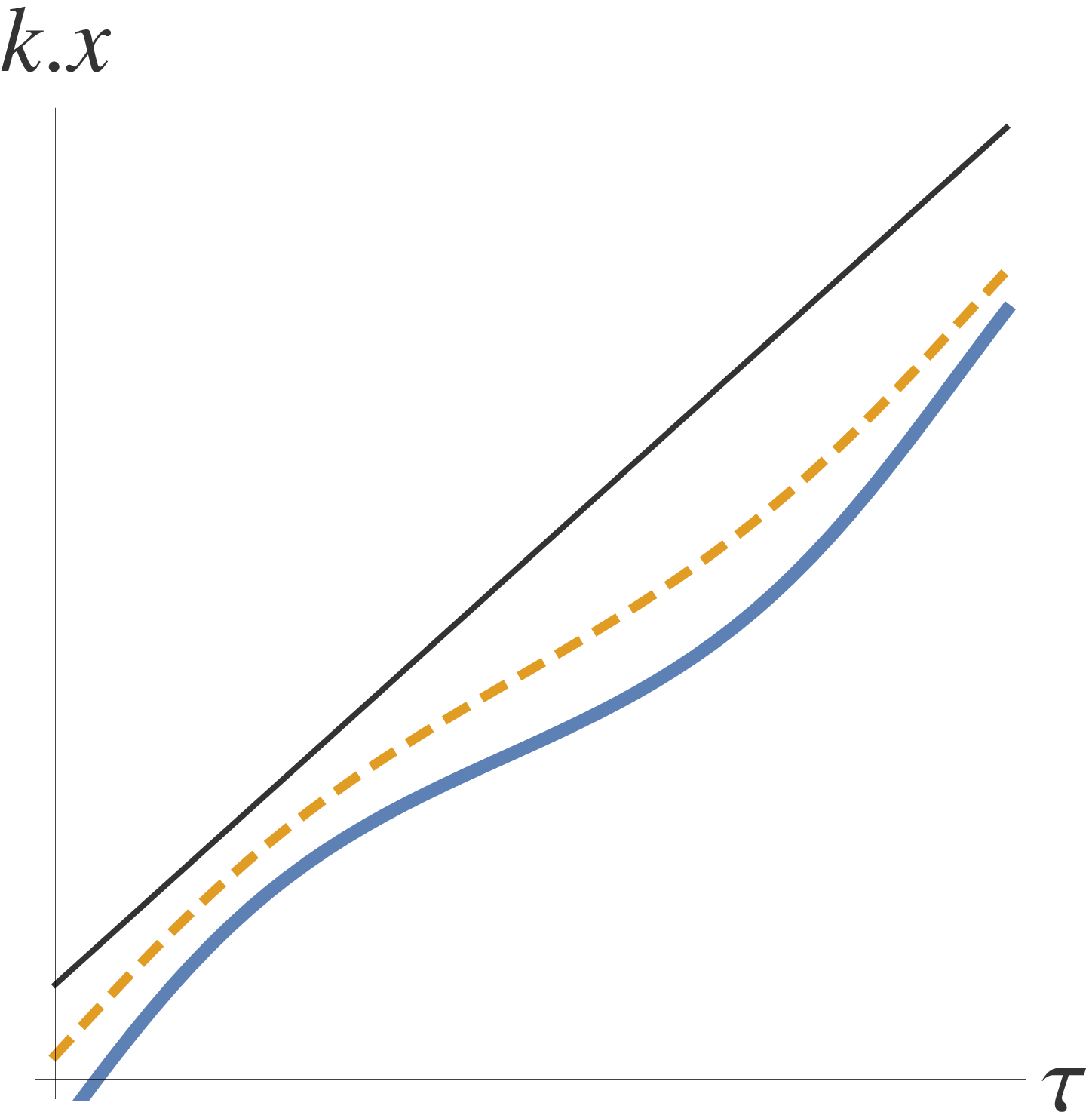}
 \caption{Demonstration of particle 
orbits at the magnetic node of a standing plane wave of $\xi_{1}=\xi_{2}=10$, 
$\omega_{1}=\omega_{2}=0.01m$, $\kbar^{2}=(0.01\,m)^{2}$. In order of increasing 
parameter $s=\mu_{\perp}^{2}/\varpi_{\perp}^{2}$: \emph{Thin solid line:} $p_{\tin}=-a_{\tin}$ i.e. 
$\Pi_{\tin} = 0$ and $s=0$; \emph{Dashed line:} $p_{\tin}\cdot \eps_{1}= a_{\tin}\cdot \eps_{1}$; 
$p_{\tin}\cdot \eps_{2}= 0.2 a_{\tin}\cdot \eps_{2}$, $s=0.72$; \emph{Thick solid line:} $p_{\tin}\cdot 
\eps_{1}\to 0$; $p_{\tin}\cdot 
\eps_{2}\to 0$, $s=0.96$. (In the right-hand plot $k=\kbar$.) As
$s \to 1$, the trajectory tends to a straight line and the phase 
tends to a horizontal asymptote.}
 \label{fig:perptraj} 
\end{figure}
Example trajectories are plotted in \figref{fig:perptraj}. When $\Pi_{\tin} = 0$, $s=\mu_{\perp}^{2}/\varpi_{\perp}^{2}=0$, and the well-known circular trajectory in a circularly-polarised background is recovered (e.g. as analysed in \cite{elkina14}). As $s\to 1$, the trajectory tends towards a straight line and $\lim_{\tau\to\infty}\vphib(\tau) \to \textrm{const}$. At a magnetic node, since $\xi_{1}=\xi_{2}$, $z=0$ and $p_{z}=0$ are required for the solution to be stable, there is no ``plane wave'' limit to compare against.
\newline

\emph{Zero initial transverse canonical momentum} also leads to the PDE in \eqnref{eqn:univ1} being reducible to an ODE. The term $a^{2}$ depends just on $\vphi_{\Delta}$, whereas $p\cdot a$ is a function of both $\vphi_{\Delta}$ and $\vphi_{\Sigma}$. Using \eqnref{eqn:ptransid} and setting the initial transverse canonical momentum to zero removes all terms linear in $a$ so only $\vphi_{\Delta}$-dependent terms remain. Then 
\[
 \dot{\vphi}_{\Delta}^{2} - \dot{\vphi}_{\Sigma}^{2} = 
\frac{(\kd \cdot \pin)^{2}}{m^{2}}-\frac{(\ks\cdot \pin)^{2}}{m^{2}} -\frac{k_{\Delta}^{2}~a^{2}}{m^{2}}
\]

For this set-up of fields, the right-hand side is additively separable, giving the 
two equations:
\bea
 \dot{\vphi}_{\Delta}^{2} = \frac{(\kd\cdot \pin)^{2}}{m^{2}}  + k_{\Delta}^{2}
\left[\xis^{2} - 4\xi_{1}\xi_{2}\sin^{2}\frac{\vphi_{\Delta}}{2}\right]  \label{eqn:vphid}
\eea
\[
 \dot{\vphi}_{\Sigma}^{2} = \frac{(\ks\cdot \pin)^{2}}{m^{2}}.
\]
Here we notice the similarity with the magnetic node case \eqnref{eqn:perptrajeqn1}. It is straightforward to show these yield the solution:
\bea
 \vphid = 2\,\am\left( 
\frac{\varpi_{\Delta}\tau}{2} 
\Bigg| -\frac{\mu_{\Delta}^{2}}{\varpi_{\Delta}^{2}}  \right); \qquad \vphis = \frac{\ks\cdot \pin}{m}\, \tau \label{eqn:vphid1}
\eea
\[
 \varpi_{\Delta}^{2} = \frac{(\kd \cdot \pin)^{2}}{m^{2}}+k_{\Delta}^{2}\xis^{2}; \qquad 
\mu_{\Delta}^{2} = -4 \kd^{2}\xi_{1}\xi_{2}, \checkedb
\]
where $\tau$ is measured from 
$\vphid=\vphis = 0$. (Also in this case an analytical expression for the transverse co-ordinates of 
the electron's trajectory can be ascertained by solving \eqnref{eqn:class0}, but the solution in 
terms of Jacobi and elliptic functions is again unilluminating.) The signs of the square roots were 
chosen so that in the plane-wave limit
$k_{2}\to k_{1}$, $a_{2}\to 0$, one recovers the result $\vphi_{j} = (k_{j} \cdot \pin/m) \tau$ (see 
e.g. \cite{ilderton09}). This can be seen directly when taking the plane-wave limit of the solution 
\eqnref{eqn:vphid1}, for which $\mu_{\Delta}/\varpi_{\Delta} \to 0$ and $\trm{am}(x|0) = x$. As 
$|\mu_{\Delta}/\varpi_{\Delta}|$ varies between permitted values $0\leq 
|\mu_{\Delta}/\varpi_{\Delta}| < 1$, it therefore interpolates between the plane-wave limit and what 
one could call the \emph{standing-wave limit}. We highlight that the classical solution predicts 
forbidden parameter regions for the particle dynamics. If $\kd^{2}a^{2}>(\kd\cdot \pin)^{2}$, there 
is no \emph{real} solution for $\dot{\vphi}_{\Delta}$ (in \eqnref{eqn:vphid}) and therefore for 
the phase. We will see this condition reappear in the quantum treatment.
\newline

\begin{figure}[!h] 
\centering
\includegraphics[draft=false,width=4cm]{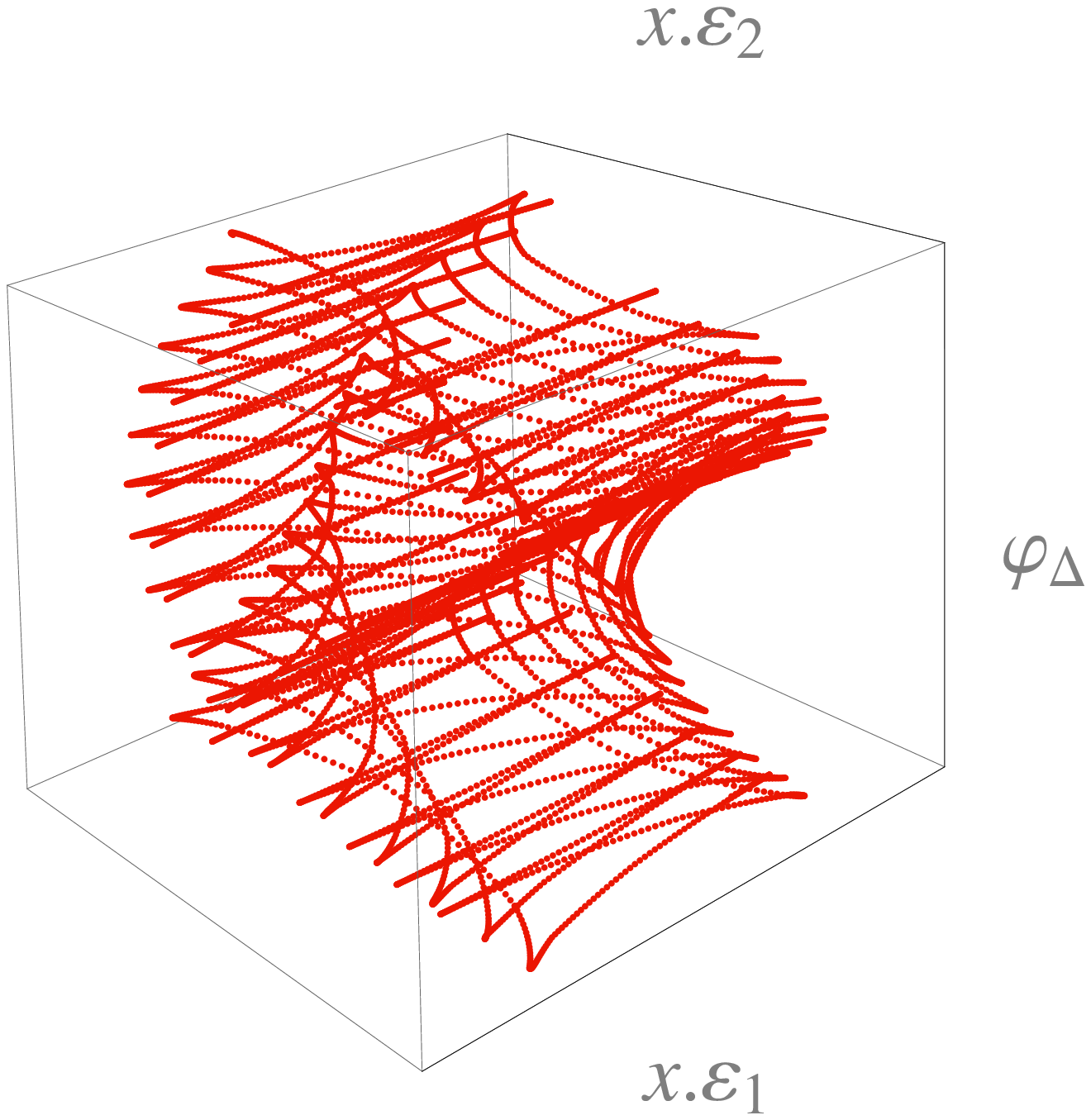}\hfill
\includegraphics[draft=false,width=4cm]{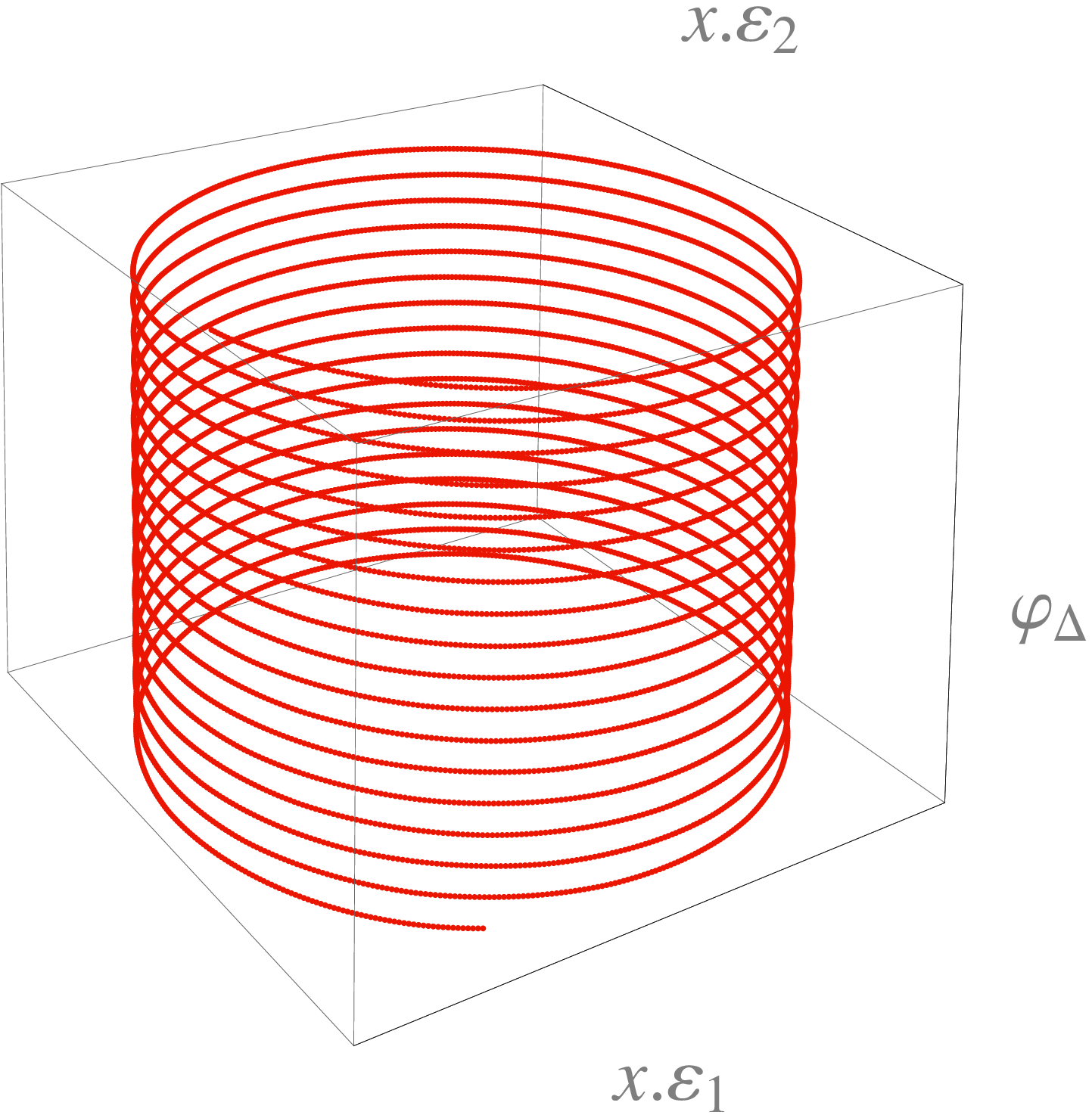}
 \caption{Demonstration of different particle orbits in a plane wave for $p_{\tin}=m(20.0255,0,0,20.0005)$, $\omega_{1}=\omega_{2}=0.01\,m$, $-\kd^{2}=(0.02\,m)^{2}$. \emph{Left}: $\xi_{1}=\xi_{2}=10$, \emph{Right}: $\xi_{1}=0.01$, $\xi_{2}=10$.}
 \label{fig:classtraj} 
\end{figure}

An example of how the standing-wave limit compares, is given in  
\figref{fig:classtraj}. The standing-wave limit shows a trajectory with cusps forming a twisted 
helical structure, which is compared to the particle trajectory when $\xi_{1} \ll \xi_{2}$. 
\cole{Even though $-\kd^{2} \ll m^{2}$, we see different dynamics from that in a plane wave. This 
follows as the solution parameter $|\mu_{\Delta}/\varpi_{\Delta}| \to 1$ if the limit $\kd\cdot 
\pin\to 0$ is taken. Although the circular trajectory with longitudinal drift is expected in a plane 
wave, and is seen in \figref{fig:classtraj}, this is not actually the plane-wave limit because 
$k_{2}\neq k_{1}$. Instead, we label this the modified-plane-wave limit, which is reached when 
$(\kd\cdot \pin)^{2}\gg \kd^{2}a^{2}$, i.e. it corresponds to \emph{high energies}.}
A further type of dynamics occurs when $\xi_{1}=\xi_{2}$ and $\omega_{1}=\omega_{2}$ at 
$\vphid = 2n\pi$, $n\in \mathbb{Z}$, the particle is at a magnetic node (electric antinode), and 
will remain at rest (most 
easily seen from the 
equation for $\dot{\vphi}_{\Delta}$).

\section{Quantum Dynamics}
Here we solve the KG equation for a charged particle in the same (classical) electromagnetic 
background as in the previous section \eqnref{eqn:a1}, and will consider the same two soluble 
cases. The KG equation can be written
\[
\left[D^{2}+m^{2}\right]\Phi = 0; \qquad D = \partial + ia.
\]
Beginning with a Volkov-like ansatz: $\Phi = w(\vphi_{1},\vphi_{2})\exp(i p\cdot x)$, one acquires:
\bea
 && 2 k_{1}\cdot k_{2}\,w''_{12} +2i\,\Pi\cdot 
 (k_{1}w_{1}' + k_{2}w_{2}')  \nn \\
&& - (\Pi^{2}-m^{2})w = 0, \label{eqn:kg1}
\eea
where subscript $1$,$2$ correspond to differentiation by $\vphi_{1}$ and $\vphi_{2}$ respectively. 
In the plane-wave limit, $k_{2} \to k_{1}$, $a_{2}\to 0$, then $k_{1}\cdot k_{2} = 0$ and we 
recover the Volkov solution:
\[
\Phi(\vphi) = \exp\left\{i \left[p\cdot x + u^{\trm{pw}}(\vphi)\right]\right\},
\]
where we define the \emph{Volkov exponent}:
\bea
u^{\trm{pw}}(\vphi) = -\int^{\vphi} \frac{2 p \cdot a(\phi) +a^{2}(\phi)}{2 k\cdot p}\,d\phi. 
\label{eqn:u}
\eea
\colbk{In the quantum treatment, which relies upon asymptotic states, $p$ is the incoming particle 
momentum and as such can be identified with $\pin$ from the previous section on classical dynamics.}
\newline

\emph{Transverse motion at a magnetic node} can be found when searching for a solution 
$\vphi_{1}=\vphi_{2}=\vphib = \kbar \cdot x$ and inserting an ansatz $\Phi=w(\vphib)\exp(ip\cdot x)$ 
into the KG equation. Then instead of the PDE \eqnref{eqn:kg1}, one acquires an ODE in $\vphib$:
\bea
 && \kbar^{2}\,w'' +2i\,p\cdot \kbar~w' +\left[\xi_{\Sigma}^{2}- 2a\cdot p\right]w = 0. \label{eqn:kg1c} \checkedb
\eea
When compared to the KG in a non-lightlike plane wave (e.g. Eq. (16) in \cite{king16b} ~\footnote{In this paper, a different convention for the Volkov ansatz $\exp(+i p\cdot x)$ is used.}), we see that the KG equation at a magnetic node is equivalent in form to a plane wave with timelike wavevector ($\kbar^{2}>0$), i.e. in an \emph{electric vacuum}. (In a homogeneous, time-dependent electric field, a variety of phenomena such as pair-creation and Cerenkov radiation are expected to occur \cite{becker77}.) Putting the equation in normal form using:
\[
w = F(\vphib)\,\exp\left[-i\,\frac{ \kbar\cdot p}{\kbar^{2}}~\vphib\right], \checkedb
\]
one acquires:
\bea
\frac{d^{2}F(y)}{dy^{2}} + \left[\lambda - 2Q \cos(2y)\right]F(y) = 0, \label{eqn:Mat0}
\eea
\bea
\lambda = \lambda_{\perp} = 4~\frac{(\kbar\cdot p)^{2}+ \xi_{\Sigma}^{2}m^{2}\kbar^{2}}{\kbar^{4}}; \quad Q = Q_{\perp} = 4~\frac{\xi_{\Sigma}m|p^{\perp}|}{\kbar^{2}} \nn\\ \checkedb
\eea
and $y = (\vphib-\vphi_{0}-\pi/4)/2$.
\eqnref{eqn:Mat0} is a canonical form of the Mathieu equation \cite{nist_dlmf}. In the current situation, it resembles the Schr\"odinger equation for a sinusoidal potential. We discuss the solutions of the Mathieu equation after presenting the solution to the second case.
\newline

Again, one can question the stability of the magnetic node solution, especially since one is dealing with a wavefunction in the quantum case and not a point particle as in the classical case. To answer this, consider the current
\[
j^{\mu} = \Phi^{\dagger}\partial^{\mu}\Phi - \partial^{\mu}\!\left(\Phi^{\dagger}\right)\Phi + 
 2 a^{\mu} \Phi^{\dagger}\Phi,
\]
where $\Phi=\Phi(\vphi_{1},\vphi_{2})$. What concerns us is the longitudinal current. If we use $\Phi=w(\vphi_{1},\vphi_{2})\exp(i\,p\cdot x)$, then:
\bea
j^3(\vphi_{1},\vphi_{2}) 
&=& -2ip^{3} |w|^{2} + k_{1}^{3}\left(w^\ast w_{1}' - w_{1}^{\ast\,\prime} w\right) \nn \\
&& + 
k_{2}^{3}\left(w^\ast w_{2}' - w_{2}^{\ast\,\prime} w\right).
\eea
We see that if $p^{3}=0$ and $k_{1}^{3}=-k_{2}^{3}$ then $j^{3}$ is both antisymmetric in exchange of arguments, but necessarily symmetric in the limit $\vphi_{1}-\vphi_{2}\to 0$, so we can conclude $j^{3}$ is identically zero. Therefore at this order of calculation, also in the quantum case the magnetic node solution is stable.
\newline

\emph{Zero initial transverse canonical momentum} can be solved for in the quantum case, by making the ansatz $\Phi = w(\vphid)\exp(i p\cdot x)$ to rewrite the KG equation as:
\bea
 && k_{\Delta}^{2}\,w''+2i\,\Pi\cdot 
 k_{\Delta}w'  - (\Pi^{2}-m^{2})w = 0. \label{eqn:kg2}
\eea
We recall that $p$ is the asymptotic viz. the initial free-particle momentum, assuming the external 
field is switched on adiabatically. Compared with the classical case, a simplification is acquired 
in the quantum case already if $p_{\tin}^{\perp}=0$. Then if the field is switched on 
adiabatically, also $a_{\tin}^{\perp}=0$, so the condition on $p_{\tin}^{\perp}$ is
equivalent to requiring $\Pi_{\tin}^{\perp}=0$.
In this case the canonical momentum squared is again additively separable in $\vphid$ and 
$\vphis$. Because the choice of background gives no dependency on $\vphis$, it is not required in 
order to parametrise the wavefunction. The $\vphi_{\Delta}$-dependent terms then give:
\[
 k_{\Delta}^{2}~w'' + 2i p \cdot \kd~w' -  a^2 w = 0.
\]
This is reminiscent of the KG equation in a non-lightlike plane wave but with the phase variable 
replaced with $\vphid$, where since $k_{\Delta}^{2} < 0$, the plane wave is in a \emph{magnetic 
vacuum}, a situation recently analysed in \cite{king16b}. We then apply the simple 
transformation:
\bea
w=J(\vphid)\exp\left[-i~\frac{\kd \cdot p}{k_{\Delta}^{2}}\,\vphid\right], \label{eqn:ww1} \checkedb
\eea
to acquire:
\bea
J'' + \frac{1}{k_{\Delta}^{2}}\left[\varpi_{\Delta}^{2} + \mu_{\Delta}^{2}\sin^{2}\frac{\vphid}{2}\right]J = 
0  .
\label{eqn:Mat1}
\eea

This is again the Mathieu equation \eqnref{eqn:Mat0} with:
\[
\lambda = \lambda_{\Delta} = \frac{4[(\kd\cdot p)^{2}+(\xi_{1}^{2}+\xi_{2}^{2})m^{2}k_{\Delta}^{2}]}{k_{\Delta}^{4}};
\]
\[ 
Q = Q_{\Delta} = -\frac{4\xi_{1}\xi_{2}}{k_{\Delta}^{2}},
\]
and $y = \varphi_{\Delta}/2$. Integration constants have once again been chosen to reproduce 
the correct zero-field limit. The plane-wave limit is 
acquired by taking the limit $k^{2}_{\Delta} \to 0$ and $a_{2}\to 0$. \eqnref{eqn:Mat1} then tends to the potential-free Schr\"odinger equation and 
the $\vphid$-dependent terms in the phase become:
\bea
\frac{\vphid}{k_{\Delta}^{2}}\left[-\kd\cdot p + 
\sqrt{(\kd\cdot p)^{2}-k_{\Delta}^{2}a^{2}}\right] \to \frac{-\vphid 
~a^{2}}{2\kd\cdot p}. \label{eqn:pwl1b}
\eea
Supposing $k_{2}=k_{1}(1-\delta)$, and taking the limit $\delta\to 0$, the plane-wave (Volkov) 
exponent \eqnref{eqn:u} is recovered (recall, $a\cdot p =0$).
\newline

\begin{figure}[!h] 
\centering
\includegraphics[draft=false,width=6cm]{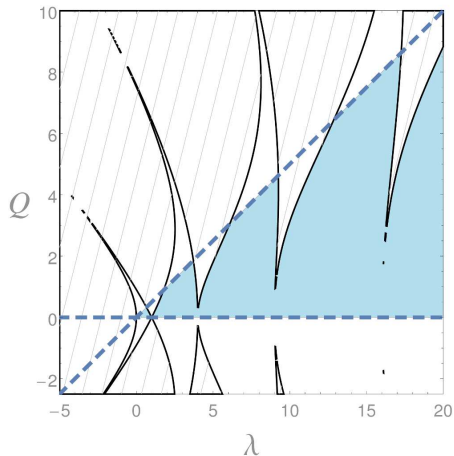}$\qquad$
 \caption{Regions of instability (gaps) in $\lambda$-$Q$ space where $\trm{Im}\,\nu \neq 0$, are indicated by the linear hatched regions. In both the magnetic-node and zero initial transverse canonical momentum cases, allowed parameters are in the shaded region between the dashed lines, $0\leq Q<\lambda/2$. There are no stable states for $\lambda<0$, which corresponds to the classically-forbidden region. }
 \label{fig:floquet1} 
\end{figure}

The Mathieu equation is common to both electric and magnetic cases. The solution to the Mathieu 
equation can be written in terms of the \emph{Mathieu characteristic exponent} or \emph{Floquet 
exponent} $\nu(\lambda,Q)$ \cite{muellerkirsten06}:
\bea
 J(\vphid) = \e^{i\nu\vphid/2}\,\phi(\vphid/2); \quad \phi(\vphid/2+2 n \pi) = \phi(\vphid/2), \nn \\\label{eqn:floq1}
\eea
which depending on the sign of its imaginary part, can represent stable regions (bands) 
($\trm{Im}\,\nu = 0$) or unstable regions (gaps), in which the wavefunction diverges 
(plotted in \figref{fig:floquet1}). The only physical solution for the wavefunction in the gaps is 
the trivial solution $J=0$. When the coupling $Q$ is \emph{weak}, the gaps 
become narrower and in the limit $Q\to 0$, they become infinitesimally thin and parameter values 
become continuous. This occurs at high particle energy where the dynamics approach the plane wave 
limit. When the coupling $Q$ increases, so do the widths of the gaps and tunnelling between bands 
becomes increasingly suppressed \cite{muellerkirsten06}. Just as in the classical case, we see from 
\figref{fig:floquet1} that, 
for the case of zero initial transverse canonical momentum, also in the quantum case no solution 
exists for when $(\kd\cdot p)^{2} <  \kd^{2}a^{2}$, where $\lambda_{\Delta} <0$.
%
%
%
\section{Approximations}
The solution to the Mathieu equations gives the exact wavefunction for the two cases under consideration. However, it is useful to study how these wavefunctions can be approximated so that scattering calculations become practicable. In particular, how the solutions compare with using the plane wave model. Approximations to solutions of the Mathieu equation have recently been discussed in the context of a magnetic vacuum \cite{king16b} and we reiterate some of the arguments here in terms of the magnetic vacuum case of zero initial transverse canonical momentum in a standing wave.
\newline

Rather than studying bands and gaps, an alternative perspective is provided by returning to the 
Mathieu equation for $J$ in \eqnref{eqn:Mat1} and recognising that, \colH{generally speaking for 
intense optical laser fields and \colb{initially accelerated particles}}, $\kd^{2}/m^{2}$ is the 
smallest parameter. Since the smallest parameter multiplies the highest derivative, the problem is 
well-suited to multiple scale perturbation theory \cite{bender78}. When applied to the 
Schr\"odinger equation, the leading order approximation is equivalent to the leading-order 
approximation from WKB \cite{jeffreys25}. Here, we use multiple scale analysis 
(singular perturbation theory) on \eqnref{eqn:Mat1}, which can be viewed as the equation of a 
simple harmonic oscillator with a slowly-varying natural frequency. The multiple scale approach 
consists of defining another propagation scale $\vphi^{\trm{\tiny{ms}}}_{\Delta} = f(\vphid)$ and 
demanding that the natural frequency of the oscillator be 
constant \cite{bender78}. For \eqnref{eqn:Mat1}, this new propagation scale (whose sign was again chosen to reproduce the 
free-particle wavefunction in the zero-field limit) is:
\bea
 \vphi^{\trm{\tiny{ms}}}_{\Delta}(\vphi_{\Delta}) =  \frac{m}{k_{\Delta}^{2}}\,\int^{\vphid}\! d\phi ~ \dot{\vphi}^{\trm{\tiny{cl}}}_{\Delta}(\phi) \label{eqn:phims1}
\eea
\bea
\dot{\vphi}^{\trm{\tiny{cl}}}_{\Delta}(\phi) =\sqrt{\frac{(\kd\cdot p)^{2}}{m^{2}}  + k_{\Delta}^{2}
\left[\xi_{\Sigma}^{2} - 4\xi_{1}\xi_{2}\sin^{2}\!\frac{\phi}{2}\right]}. \label{eqn:T}
\eea

Comparing this with the classical case \colH{\eqnref{eqn:vphid}}, we see $\vphi^{\trm{\tiny{ms}}}_{\Delta}$ is just the 
integral of the classical $\dot{\vphi}_{\Delta}$, which we denote 
$\dot{\vphi}^{\trm{\tiny{cl}}}_{\Delta}$, with respect to the ``timescale'' 
$-m \,k\cdot x/k_{\Delta}^{2}$. Identifying $k_{\Delta}^{2}$ with $\hbar$ and introducing the 
particle 
energy $\mathcal{E}$, the relation of the new scale $ \vphi^{\trm{\tiny{ms}}}_{\Delta}$ to the classical quantity $\vphi_{\Delta}^{\trm{\tiny{cl}}}$ is reminiscent of the relation between the quantum phase $i\mathcal{E}t/\hbar$ and the classical phase $i\omega t$.
\newline

Including the first three terms in the multi scale perturbation expansion gives:
\bea
 J^{\trm{\tiny{ms}}}(\vphi_{\Delta}) =
\left(1-\frac{k_{\Delta}^{2}\,a^{2}(\vphid)}{(\kd\cdot p)^{2}}\right)^{-1/4}\Bigg|_{O(k_{\Delta}^{2})}\exp\left[i u^{\trm{ms}}(\vphid)\right],\nn\\\label{eqn:MSPT1}
\eea
where in line with the notation of other exponents in this paper, we define
$u^{\trm{ms}}(\vphid) = \vphi^{\trm{ms}}_{\Delta}(\vphi_{\Delta})$.
Indeed, by including higher orders in the singular perturbation expansion, the band-like structure 
of the Mathieu solution can be reconstructed \cite{bender78}, so one might expect this approximation 
to include all physical effects. As already pointed out in \eqnref{eqn:pwl1b}, if $(\kd\cdot 
p)^{2}\gg \kd^{2}a^{2}$ and the square-root is expanded, then the modified-plane-wave limit is 
recovered. In contrast, if the electron starts from rest and the 
frequencies of the two plane waves are equal, then $\kd\cdot p = 0$, \colb{and the plane wave model 
is not applicable}. \colb{Therefore, we should suspect the plane wave model to become questionable 
at some point between these two situations, when $(\kd\cdot p)^{2}\not\gg \kd^{2}a^{2}$}. 
\newline

The most drastic but versatile approximation is what we call the \emph{high energy approximation}. 
By using the following product ansatz in the original KG equation:
\bea
 w = F(\vphi_{1})G(\vphi_{2})H(\vphid)\e^{i p\cdot x}, \label{eqn:HEAansatz1}
\eea
 one acquires:
\bea
k_{\Delta}^{2} \left[\frac{H''}{H} - \frac{1}{2} \frac{F'}{F}\frac{G'}{G} +\frac{1}{2}\frac{F'}{F}\frac{H'}{H}  -\frac{1}{2} \frac{G'}{G}\frac{H'}{H}\right] + 2i 
\kd\cdot p~\frac{H'}{H}  \nn && \\ +   2 i k_{1}\cdot p \frac{F'}{F} + 2 i k_{2}\cdot p
\frac{G'}{G} - 2p\cdot a_{1} - 2p\cdot a_{2}- a \cdot a
= 0.\nn \\\checked\label{eqn:G}
\eea
Neglecting all terms of order $\kd^{2}$, makes the equation additively separable and the 
field-dependent terms can be exponentiated:
\bea
H &=& \e^{-i\int^{\vphid} \frac{a^{2}}{2 \kd\cdot p}}; \quad F = 
\e^{-i\int^{\vphi_{1}} \frac{p\cdot a_{1}}{k_{1}\cdot p}}; \quad G = 
\e^{-i\int^{\vphi_{2}} \frac{p\cdot a_{2}}{k_{2}\cdot p}}.\nn \\\label{eqn:H1}&& \checked
\eea
We call this approach that neglects second-order derivatives in the KG equation in this way the 
``high energy approximation''. (\eqnref{eqn:H1} is the 
``simplified solution'' presented in a recent analysis of the Klein-Gordon
equation \colb{(Eq. 76 of \cite{hu15})} for a related but different case to the one studied here, 
of a high-energy particle colliding obliquely with counter-propagating laser waves.) \colbkq{For 
the zero initial transverse canonical momentum case, 
this approximation can be related to the more accurate multi scale solution, which \emph{does} 
take into 
account the second derivative.} To justify the labelling ``high energy approximation'', 
we 
recall arguments from \cite{king16b} for the convenience of the reader. 
Ostensibly, one might presume that this high energy approximation is valid when $\kd^{2}$ is the 
smallest parameter so the second derivative can be neglected. However, when this is the case, it is 
multiplying the largest derivative, and so 
when the approximation is made, one is assuming ``$\kd^{2} \times \trm{quadratic
derivatives}$'' 
is the smallest term in \eqnref{eqn:G}. Moreover, when the high energy approximation is used, one is 
assuming that the solution is perturbative in $\kd^{2}$ and one is calculating the leading-order 
term. However \eqnref{eqn:G} cannot be attacked using regular perturbation theory because, when 
$\kd^{2} \to 0$, one of the solutions disappears \cite{bender78}. Instead, it is of the form that 
\emph{singular perturbation theory} may work, in other words when an asymptotic approximation is 
useful. We refer to \eqnref{eqn:H1} as the high energy approximation because, when one expands the 
square root of the singular perturbation result \eqnref{eqn:T} for $(\kd\cdot p)^{2} \gg 
\kd^{2}a^{2}$ and combines it with the rest of the solution in \eqnref{eqn:ww1}, one acquires the 
approximate solution of the KG equation:
\[
w \approx H(\vphid)\e^{i p\cdot x}.
\]
For the case of zero initial transverse canonical momentum, since we implicity assume that $a$ is 
zero in the infinite past and future, $p\cdot \eps_{1} = p\cdot \eps_{2} = 0$, so $F$ and $G$ are 
unity in this case. Then, we see that  $(\kd\cdot p)^{2} \gg \kd^{2}a^{2}$ corresponds to 
\eqnref{eqn:HEAansatz1}, justifying the term ``high energy approximation''. Therefore the condition 
$\kd^{2}$ being the smallest parameter is not sufficient to acquire \eqnref{eqn:H1}. If the plane 
waves are counter-propagating and of equal 
frequency then the condition $(\kd\cdot p)^{2} \gg \kd^{2}a^{2}$  reduces to $(p_{z}/m)^{2} 
\gg \xi_{\Sigma}^{2}$, for longitudinal particle momentum $p_{z}$, which is very 
similar to the requirement $\gamma \gg \xi$ in recent approaches to derive electron states 
\cite{dipiazza14} and propagators \cite{dipiazza15} of ultrarelativistic electrons in general 
background fields.
\newline

We will begin the following section with the high energy approximation of the magnetic node 
solution. \colbkq{Unlike for the zero initial transverse canonincal momentum case, we will see 
there is no condition for the high energy approximation to be valid at the magnetic node, and it 
cannot be related to more accurate approximations.} \colbkq{Na\"ively, the condition for the high 
energy approximation to be valid would be $(\kbar\cdot p)^{2} \gg -\kbar^{2}(a^{2}+2a\cdot p)$, 
since here $p\cdot a \neq 0$}. \colbk{The high energy approximation can be acquired by taking the 
limit $\vphid \to 0$ in \eqnref{eqn:H1}, which gives:
\[
\Phi(\vphib) = \exp\left\{i \left[p\cdot x + u^{\trm{pw}}(\vphib)\right]\right\}.
\]
(This solution can also be acquired by solving \eqnref{eqn:kg1c} in the limit $\kbar^{2} \to 0$). The plane wave limit is then acquired when $\vphib \to \vphi$ and $a_{2}\to 0$.}
%
%
%
\section{Photon emission}
In this section we consider some aspects of photon emission (nonlinear Compton scattering) in the 
magnetic node solution to the KG equation. First, let us study the situation in (3+1)D. We 
recall the interaction Lagrangian density in scalar QED (sQED) can be written 
\cite{itzykson80}:
\[
\mathcal{L}_{\tiny\trm{int}} = -i\,\bar{a}^{\mu}\Phi^{\dagger}\partial_{\mu}\Phi 
+i\bar{a}^{\mu}\partial_{\mu}(\Phi^{\dagger})\Phi +\bar{a}\cdot \bar{a} ~\Phi^{\dagger}\Phi 
\]
for the photon field $e\hat{A} = \bar{a} = \hat{a} + a$, where $a$ is the classical, external-field component discussed until now and $\hat{a}$ is the field of the emitted photon. Then we define:
\bea
\hat{a}^{\mu} = e~ \phpol^{\mu} \frac{1}{\sqrt{2Vl'^{0}}}\,\e^{i\,l'\cdot x};\quad  ~ \Phi_{p'} = 
\frac{\e^{i\,p'\cdot x+iu_{p'}(\vphib)}}{\sqrt{2Vp'^{0}}}, \label{eqn:states1}
\eea
where $\phpol^{2}=-1$ and the function $u_{p'}(\vphib)$ has yet to be chosen. Writing the 
scattering matrix element as $S_{\tiny\trm{fi}} = -\int d^{4}x\,\mathcal{L}_{\trm{int}}$, the 
calculation can proceed as usual, but with modified charged particle states. It can be shown that:
\bea
|S_{\tiny\trm{fi}}|^{2} \sim \delta^{(4)}\left(l'+q'-q-s\,k\right), \label{eqn:pcon1}
\eea
where $q$ is the quasi-momentum in the particular model used. Let us first consider the 
high energy approximation. In this case, the quasi-momentum \break $q^{\trm{he}} = p - (a^{2}/2 
p\cdot \bar{k})\,\bar{k}$, so that:
\[
(q^{\trm{he}})^{2} = m^{2}\left[1+\xis^{2} + \left(\frac{m^{2}}{\bar{k}\cdot p}\right)^{2} \frac{\bar{k}^{2}}{4m^{2}}\,\xis^{4}\right] = (m^{\trm{he}}_{\ast})^{2},
\]
is the ``effective mass'' which tends to the effective mass familiar from plane-wave calculations in 
a monochromatic circularly-polarised background \break $(q^{\trm{pw}})^{2}= m^{2}(1+\xis^{2})$ if 
the plane-wave limit $\kbar \to k$, $a_{2}\to 0$ is taken. Alternatively, one can use the multi 
scale approach for the magnetic node case to calculate the quasi-momentum. The non-trivial phase in 
this approach is of the form \eqnref{eqn:phims1}, which can be integrated analytically. Then the 
phase dependency of the KG solution is:
\[
u^{\trm{ms}}_{p}(\bar{\vphi})=-\frac{\kbar\cdot p }{\kbar^{2}}\,\vphib+\frac{\,\varpi_{\perp,p}m}{\kbar^{2}}\,\trm{E}\left(\frac{\bar{\vphi}}{2}\Bigg| \frac{\mu_{\perp,p}^{2}}{\varpi_{\perp,p}^{2}}\right),
\]
where the constants $\varpi_{\perp,p}$ and $\mu_{\perp,p}$ are taken from the classical phase \eqnref{eqn:perptrajeqn1}, but now in the quantum case with $\Pi_{\tin} = p$ so that:
\[
\varpi_{\perp,p}^{2} = \frac{(\kbar\cdot p)^{2}}{m^{2}}+\kbar^{2} \xis^{2}\left[1+ 
\frac{2|p^{\perp}|}{m\xis}\right]; \quad \mu_{\perp,p}^{2} = \frac{4\kbar^{2}\xis 
|p^{\perp}|}{m},\checkedb
\]
and $\trm{E}\left(\cdot | \cdot \right)$ is the elliptic integral of the second kind \cite{nist_dlmf}. Just as the cycle average of the Volkov exponent $u^{\trm{pw}}(\vphi)$ is separated out of the rest of the integral to acquire the quasi-momentum, so too can here the average of the multiple scale exponent be separated out:
\[
u^{\trm{ms}}_{p}(\bar{\vphi}) \approx \av{u^{\trm{ms}}_{p}}\,\bar{\vphi} + \int^{\bar{\vphi}}\Delta u,
\]
\[
\av{u^{\trm{ms}}_{p}} = -\frac{\kbar\cdot p}{\kbar^{2}}+\frac{2}{\pi}\frac{\varpi_{\perp,p}m}{\kbar^{2} }\,\trm{E}\left(\frac{\pi}{2}\Bigg| \frac{\mu_{\perp,p}^{2}}{\varpi_{\perp,p}^{2}}\right) 
\]
where the cycle-average of the remaining term $\Delta u$ is zero (a study of the dependence of this 
type of approximation on pulse duration can be found in \cite{harvey12}). Then the quasi-momentum 
becomes $q^{\trm{ms}}= p+\av{u^{\trm{ms}}_{p}}\,\kbar$. One can likewise define a quasi-momentum for 
the exact solution, by using the Mathieu characteristic exponent from \eqnref{eqn:floq1} to give:
\[
q = p - \left[\frac{\kbar\cdot p}{\kbar^{2}}-\frac{\nu_{\perp,p}(\lambda_{\perp},Q_{\perp})}{2}\right]\,\kbar. \checkedb
\]
We recall that:
\[
\frac{Q_{\perp}}{\lambda_{\perp}} = \frac{|p^{\perp}|\xis}{1+|p^{\perp}|^{2}+\xis^{2}} \leq \frac{1}{2}.
 \] 
If $|p^{\perp}|$ is much greater or much less than $\xis$, this ratio is much less than one, and 
$\nu_{\perp,p}(\lambda_{\perp},Q_{\perp})\approx \sqrt{\lambda_{\perp}}$, \cite{muellerkirsten06}, 
which immediately gives the connection with the multi scale approach \eqnref{eqn:T}. The accuracy 
of this approximation therefore gives a condition for when the accuracy of the multi scale approach should be 
good.
\newline

\begin{figure}[h] 
\centering
\includegraphics[draft=false,width=8cm]{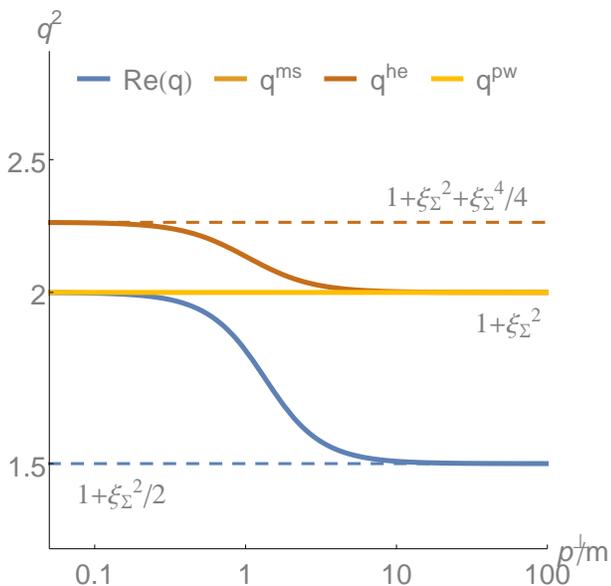}$\qquad$
 \caption{A log-log plot 
 effective mass varies with increasing transverse momentum for $\xis=1$, $\kbar^{2}=(0.01\,\trm{m})^{2}$. The multi scale result agrees exactly with the analytical result from the Mathieu characteristic exponent, whereas the high energy approximation disagrees at both high and low energy.}
 \label{fig:quasimom1} 
\end{figure}

\begin{figure}[h] 
\centering
\includegraphics[draft=false,width=8cm]{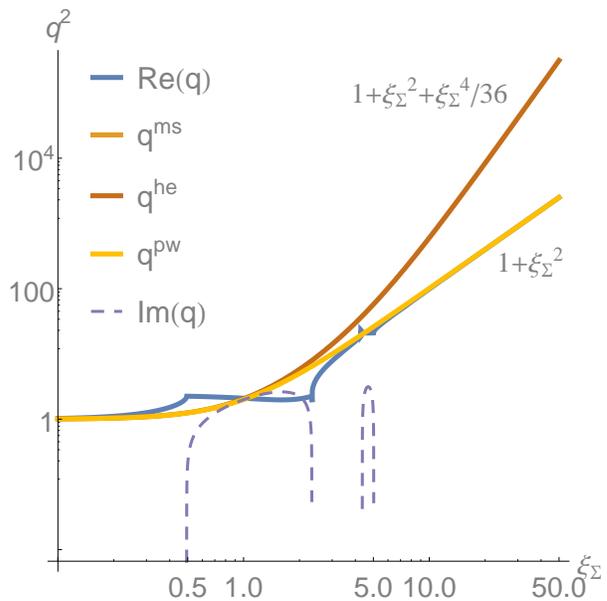}$\qquad$
 \caption{A log-log plot of how the effective mass varies with increasing intensity parameter $\xis$ 
for $\kbar^{2}=(5\,\trm{m})^{2}$ and $p^{\perp}=2\,m$. \colb{The large value of $\kbar^{2}$ has been 
chosen to emphasise the band structure. For $\kbar^{2}\ll 1$, the discrepancies between the high 
energy and plane wave approximation with the multi scale and exact results persist.}}
 \label{fig:quasimom2} 
\end{figure}

At the magnetic node, since the dimensionality of the system has been reduced, the high energy 
approximation and the wide-angle scattering limit 
of $p^{\perp}/m \to \infty$ are not independent of one another \Ben{I think your comments were 
referring to the approximation for the zero transverse canonical momentum case. Those arguments must 
be adapted to here. The idea is that $p^{0}$ cannot be large without $|p^{\perp}|$ in the plane, in 
contrast to the usual case.}. \colbk{This can be seen by considering the high energy 
approximation phase dependency, which we recall is the Volkov exponent with the wavevector replaced 
with $\kbar$:
\[
 u^{\trm{he}}(\vphib) = -\int^{\vphib} \frac{2 p \cdot a(\phi) +a^{2}(\phi)}{2 
\kbar\cdot p}\,d\phi. \label{eqn:uhe} 
 \]
 Now, in the usual plane-wave case, the limits $k\cdot p \to \infty$ and $p\cdot a \to \infty$ are 
independent of one 
another. Here however, since $\kbar\cdot p = \omega [m^{2}+(p^{\perp})^{2}]^{1/2}$, and
$p\cdot a = \xi|p^{\perp}|\cos(\vphib - \vphib_{0})$ where $\tan\vphib_{0} = p\cdot 
\eps_{2}/p\cdot \eps_{1}$, the high energy limit and the wide-angle scattering limit, 
which, in a plane wave correspond to different physics, are at the magnetic node of a standing wave, connected.} A consequence 
is that the high energy 
approximation, where one expects the multi scale result to agree with the plane-wave limit, is 
also the wide-angle scattering limit and hence the approximation of neglecting the second 
derivative term in the KG 
equation becomes \emph{worse}, not better, as displayed in \figref{fig:quasimom1}.
Since the high energy approximation is (n\"aively) expected to be useful when $(\kbar\cdot p)^{2} 
\gg -\kbar^{2}(a^{2}+p\cdot a)$, it is not surprising that for strong fields, the high energy 
effective mass $(q^{\trm{he}})^{2}$, diverges from the exact result. More surprising is that for 
strong fields, the multi scale and exact results tend to the plane wave limit, as displayed in 
\figref{fig:quasimom2} (this can be proven from \eqnref{eqn:kg1c}). None of the approximations 
capture the band structure of the exact solution, which is displayed by regions of non-zero 
imaginary quasi-momentum.

\colb{These results can be compared to the reasoning of the plane wave model. The standard argument 
\cite{ritus85} is that QED is a relativistic theory and so all observables are to be built from 
relativistic invariants. For a single seed particle of momentum $p$, four relativistic invariants 
are identified $\xi$ and $\eta = k\cdot p/m^{2}$, for a relevant external-field wavevector $k$, 
$\mathcal{F} = -e^{2}F^{\mu\nu}F_{\mu\nu}/4 m^{4}$ and $\mathcal{G} = 
-e^{2}F^{\mu\nu}F^{\ast}_{\mu\nu}/4 m^{4}$, where $F^{\ast}_{\mu\nu}$ is the dual Faraday tensor 
\cite{jackson75}. The probability of a QED process can then be expressed as a function of these 
parameters $P=P(\xi,\eta,\mathcal{F},\mathcal{G})$ and when $\mathcal{F},\mathcal{G} \ll \xi,\eta,1$ 
this can be expanded in a Taylor series in $\mathcal{F}$ and $\mathcal{G}$, the leading order of 
which is $P(\xi,\eta,0,0)$ \footnote{In the usual discussion of the validity of the plane wave model 
$\eta$ is replaced with $\chi = \xi \eta$, but the reasoning is the same.}. Assuming the depedency 
on $\mathcal{F}$ and $\mathcal{G}$ is perturbative, or the non-perturbative dependency is 
vanishingly small, this leading order term, which is the probability of the process in a plane wave 
background, is a valid approximation. In the magnetic node case, the relevant wavevector is $\kbar$ 
and there is a fifth relativistic invariant, $\kbar^{2}$. The (n\"aive) condition that the high energy 
approximation of Volkov (plane-wave) form is a good approximation was $(\kbar \cdot p)^{2} \gg 
-\kbar^{2}(a^{2}+p\cdot a)$, which in these invariants becomes $\eta^{2} \gg \mathcal{F}$. We can 
conclude that 
i) for the typical case of $\eta < 1$, this is a more stringent condition on the smallness of 
$\mathcal{F}$ than is usually argued, for the plane wave model to be valid and ii) where one expects 
the plane wave model to \colbkq{tend to the exact result at high particle energies, at the magnetic 
node in a standing wave it tends to the incorrect result at the level of the quasimomentum.}}

Suppose we continue with the scattering calculation. In the plane wave limit, following the standard method \cite{landau4}, we find the probability for photon emission per unit external-field phase $W = \sum_{s}W_{s}$, which can be written in the usual way as a sum over harmonics $s$, is:
\bea
W_{s} = \frac{\alpha m^{2}}{4k\cdot p} \int_{0}^{u_{s}}\!\!\frac{du}{(1+u)^{2}}\left[  -4J_{s}^{2}\,\frac{m_{\ast}^{2}}{m^{2}} + 2\xi^{2}\left(J_{s+1}^{2}+J_{s-1}^{2}\right)\right], \label{eqn:Pcurr5}\nn\\
\eea
for $m_{\ast}^{2}=m^{2}(1+\xi^{2})$, where the Bessel functions of the first kind $J_{s}$ have arguments $z$:
\[
z = \frac{2s\xi}{\sqrt{1+\xi^{2}}}\sqrt{\frac{u}{u_{s}}\left(1-\frac{u}{u_{s}}\right)},\nn \\ \label{eqn:z1}
\]
with $u_{s} = 2s\,k\cdot p/m^{2}(1+\xi^{2})$. The sQED result \eqnref{eqn:Pcurr5} is very close to the QED version \cite{landau4}, but without spin-dependent terms (it agrees with a similar recent calculation \cite{raicher16}). 

However, if one attempts to use the high energy approximation of the magnetic node solution in the same calculation, a problem becomes immediately obvious. From the longitudinal component of \eqnref{eqn:pcon1} one notes that $q'^{\parallel} = -l'^{\parallel}$. Since a requirement of the magnetic node solution is that $q'^{\parallel}= 0$, we see that after emitting a photon, the particle is, in general, placed into a \emph{different} outgoing state, and not $\Phi_{p'}$. This leads to a contradiction, so we conclude the magnetic node solution of a scalar particle in a standing wave is \emph{unstable} due to radiation emission.

That the magnetic node solution is unstable when radiative emission is taken into account, is reminiscent of ponderomotive effects on a charged particle in an inhomogeneous background. \emph{Ponderomotive trapping} is a well-known phenomenon \cite{bauer95,kaplan05} where field gradients drive electrons into \emph{minima} of the potential, i.e. magnetic \emph{antinodes}. The force on the scalar particle depends however on the phase of the field when it is scattered. If in the correct orientation, the magnetic field may produce a restoring force on the scattered particle and drive it back to the magnetic node. This is sometimes referred to as \emph{anomalous radiative trapping} \cite{gonoskov14}.
\newline
%
%
%
\section{Conclusion}
Solutions have been presented for a scalar particle in a background formed of two 
counter-propagating plane waves. Two cases were studied: when the particle is confined to a magnetic 
node (electric vacuum) and when the particle has zero initial transverse canonical momentum 
(magnetic vacuum). Both the classical dynamics (Lorentz equation) and the quantum dynamics 
(Klein-Gordon equation) were solved analytically. Different approximations to the quantum dynamics 
were presented. First, the high energy approximation of neglecting quadratic and second-order derivatives yields a 
``modified'' Volkov (plane-wave) wavefunction. Second, an asymptotic approximation using multi scale 
perturbation theory gave a WKB-like solution that retains dependency on the second derivative. Aspects of photon 
emission (nonlinear Compton scattering) were studied using the high energy approximation of 
dynamics at a magnetic node. It was found that the magnetic node solution is in general unstable 
when radiation emission is taken into account. Moreover, whilst at the magnetic node, since motion 
is confined to a plane, the high-energy and the wide-angle scattering limit become conflated. 
Describing wide-angle scattering generally requires the second derivative, and by studying the 
particle's quasi-momentum it was found that the high-energy Volkov-like approximation disagrees 
with the multi scale and exact results at low and high energies. \colb{If standard arguments about 
when the plane wave model is valid, are used to justify approximating the background as a plane 
wave, the predicted quasimomentum disagrees with the exact result.}

For the magnetic vacuum case, forbidden parameter regions were identified in the classical and quantum dynamics. If the particle starts with zero longitudinal momentum, these forbidden regions $\kd^{2}a^{2} > (\kd\cdot p)^{2}$ are quite accessible to experiment, but are completely missed in the plane wave model.
%
%
%
\section{Acknowledgments}
B. K. acknowledges support from the Royal Society International Exchanges Scheme, the generous 
hospitality of H. H., CARDC (Chinese Aerodynamics Research and Development Center) and of SWUST 
(Southwest University of Science and Technology), Sichuan, China. \colH{H. H. acknowledges the 
financial support by the National Natural Science Foundation of China under Grant No. 11204370.}
%
%
%

\bibliography{current}

\providecommand{\noopsort}[1]{}
\begin{thebibliography}{46}
\expandafter\ifx\csname natexlab\endcsname\relax\def\natexlab#1{#1}\fi
\expandafter\ifx\csname bibnamefont\endcsname\relax
  \def\bibnamefont#1{#1}\fi
\expandafter\ifx\csname bibfnamefont\endcsname\relax
  \def\bibfnamefont#1{#1}\fi
\expandafter\ifx\csname citenamefont\endcsname\relax
  \def\citenamefont#1{#1}\fi
\expandafter\ifx\csname url\endcsname\relax
  \def\url#1{\texttt{#1}}\fi
\expandafter\ifx\csname urlprefix\endcsname\relax\def\urlprefix{URL }\fi
\providecommand{\bibinfo}[2]{#2}
\providecommand{\eprint}[2][]{\url{#2}}

\bibitem[{\citenamefont{Ritus}(1985)}]{ritus85}
\bibinfo{author}{\bibfnamefont{V.~I.} \bibnamefont{Ritus}},
  \bibinfo{journal}{J. Russ. Laser Res.} \textbf{\bibinfo{volume}{6}},
  \bibinfo{pages}{497} (\bibinfo{year}{1985}).

\bibitem[{\citenamefont{Marklund and Shukla}(2006)}]{marklund_review06}
\bibinfo{author}{\bibfnamefont{M.}~\bibnamefont{Marklund}} \bibnamefont{and}
  \bibinfo{author}{\bibfnamefont{P.~K.} \bibnamefont{Shukla}},
  \bibinfo{journal}{Rev. Mod. Phys.} \textbf{\bibinfo{volume}{78}},
  \bibinfo{pages}{591} (\bibinfo{year}{2006}).

\bibitem[{\citenamefont{{Di Piazza} et~al.}(2012)}]{dipiazza12}
\bibinfo{author}{\bibfnamefont{A.}~\bibnamefont{{Di Piazza}}}
  \bibnamefont{et~al.}, \bibinfo{journal}{Rev. Mod. Phys.}
  \textbf{\bibinfo{volume}{84}}, \bibinfo{pages}{1177} (\bibinfo{year}{2012}).

\bibitem[{\citenamefont{King and Heinzl}(2016)}]{king15a}
\bibinfo{author}{\bibfnamefont{B.}~\bibnamefont{King}} \bibnamefont{and}
  \bibinfo{author}{\bibfnamefont{T.}~\bibnamefont{Heinzl}},
  \bibinfo{journal}{High Power Laser Science and Engineering}
  \textbf{\bibinfo{volume}{4}}, \bibinfo{pages}{e5} (\bibinfo{year}{2016}),
  \eprint{hep-ph/1510.08456}.

\bibitem[{\citenamefont{Narozhny and Fedotov}(2015)}]{narozhny15}
\bibinfo{author}{\bibfnamefont{N.~B.} \bibnamefont{Narozhny}} \bibnamefont{and}
  \bibinfo{author}{\bibfnamefont{A.~M.} \bibnamefont{Fedotov}},
  \bibinfo{journal}{Contemporary Physics} \textbf{\bibinfo{volume}{56}},
  \bibinfo{pages}{249} (\bibinfo{year}{2015}).

\bibitem[{\citenamefont{Becker}(1977)}]{becker77}
\bibinfo{author}{\bibfnamefont{W.}~\bibnamefont{Becker}},
  \bibinfo{journal}{Physica A} \textbf{\bibinfo{volume}{87}},
  \bibinfo{pages}{601} (\bibinfo{year}{1977}).

\bibitem[{\citenamefont{Mendonca and Serbeto}(2011)}]{mendonca11}
\bibinfo{author}{\bibfnamefont{J.~T.} \bibnamefont{Mendonca}} \bibnamefont{and}
  \bibinfo{author}{\bibfnamefont{A.}~\bibnamefont{Serbeto}},
  \bibinfo{journal}{Phys. Rev. E} \textbf{\bibinfo{volume}{83}},
  \bibinfo{pages}{026406} (\bibinfo{year}{2011}).

\bibitem[{\citenamefont{Raicher and Eliezer}(2013)}]{raicher13}
\bibinfo{author}{\bibfnamefont{E.}~\bibnamefont{Raicher}} \bibnamefont{and}
  \bibinfo{author}{\bibfnamefont{S.}~\bibnamefont{Eliezer}},
  \bibinfo{journal}{Phys. Rev. A} \textbf{\bibinfo{volume}{88}},
  \bibinfo{pages}{022113} (\bibinfo{year}{2013}).

\bibitem[{\citenamefont{Varro}(2013)}]{varro13}
\bibinfo{author}{\bibfnamefont{S.}~\bibnamefont{Varro}},
  \bibinfo{journal}{Laser Phys. Lett.} \textbf{\bibinfo{volume}{10}},
  \bibinfo{pages}{095301} (\bibinfo{year}{2013}).

\bibitem[{\citenamefont{Varro}(2014)}]{varro14}
\bibinfo{author}{\bibfnamefont{S.}~\bibnamefont{Varro}},
  \bibinfo{journal}{Laser Phys. Lett.} \textbf{\bibinfo{volume}{11}},
  \bibinfo{pages}{016001} (\bibinfo{year}{2014}).

\bibitem[{\citenamefont{Raicher et~al.}(2015)\citenamefont{Raicher, Eliezer,
  and Zigler}}]{raicher15}
\bibinfo{author}{\bibfnamefont{E.}~\bibnamefont{Raicher}},
  \bibinfo{author}{\bibfnamefont{S.}~\bibnamefont{Eliezer}}, \bibnamefont{and}
  \bibinfo{author}{\bibfnamefont{A.}~\bibnamefont{Zigler}},
  \bibinfo{journal}{Phys. Lett. B} \textbf{\bibinfo{volume}{750}},
  \bibinfo{pages}{76} (\bibinfo{year}{2015}).

\bibitem[{\citenamefont{Cronstr{\"o}m and Noga}(1977)}]{cronstroem77}
\bibinfo{author}{\bibfnamefont{C.}~\bibnamefont{Cronstr{\"o}m}}
  \bibnamefont{and} \bibinfo{author}{\bibfnamefont{M.}~\bibnamefont{Noga}},
  \bibinfo{journal}{Phys.\ Lett.\ A} \textbf{\bibinfo{volume}{60}},
  \bibinfo{pages}{137} (\bibinfo{year}{1977}).

\bibitem[{\citenamefont{Heinzl et~al.}(2016)\citenamefont{Heinzl, Ilderton, and
  King}}]{king16b}
\bibinfo{author}{\bibfnamefont{T.}~\bibnamefont{Heinzl}},
  \bibinfo{author}{\bibfnamefont{A.}~\bibnamefont{Ilderton}}, \bibnamefont{and}
  \bibinfo{author}{\bibfnamefont{B.}~\bibnamefont{King}},
  \bibinfo{journal}{arXiv preprint arXiv:1607.07449v1}  (\bibinfo{year}{2016}).

\bibitem[{\citenamefont{Uggerh\o{}j}(2005)}]{uggerhoj5}
\bibinfo{author}{\bibfnamefont{U.~I.} \bibnamefont{Uggerh\o{}j}},
  \bibinfo{journal}{Rev. Mod. Phys.} \textbf{\bibinfo{volume}{77}},
  \bibinfo{pages}{1131} (\bibinfo{year}{2005}).

\bibitem[{\citenamefont{{Di Piazza} et~al.}(2007)\citenamefont{{Di Piazza},
  Hatsagortsyan, and Keitel}}]{dipiazza07}
\bibinfo{author}{\bibfnamefont{A.}~\bibnamefont{{Di Piazza}}},
  \bibinfo{author}{\bibfnamefont{K.~Z.} \bibnamefont{Hatsagortsyan}},
  \bibnamefont{and} \bibinfo{author}{\bibfnamefont{C.~H.}
  \bibnamefont{Keitel}}, \bibinfo{journal}{Phys. Plasmas}
  \textbf{\bibinfo{volume}{14}}, \bibinfo{pages}{032102}
  (\bibinfo{year}{2007}).

\bibitem[{\citenamefont{Harding and Lai}(2006)}]{harding06}
\bibinfo{author}{\bibfnamefont{A.~K.} \bibnamefont{Harding}} \bibnamefont{and}
  \bibinfo{author}{\bibfnamefont{D.}~\bibnamefont{Lai}}, \bibinfo{journal}{Rep.
  Prog. Phys.} \textbf{\bibinfo{volume}{69}}, \bibinfo{pages}{2631}
  (\bibinfo{year}{2006}).

\bibitem[{\citenamefont{Bulanov et~al.}(2010)\citenamefont{Bulanov, Mur,
  Narozhny, Nees, and Popov}}]{bulanov10b}
\bibinfo{author}{\bibfnamefont{S.~S.} \bibnamefont{Bulanov}},
  \bibinfo{author}{\bibfnamefont{V.~D.} \bibnamefont{Mur}},
  \bibinfo{author}{\bibfnamefont{N.~B.} \bibnamefont{Narozhny}},
  \bibinfo{author}{\bibfnamefont{J.}~\bibnamefont{Nees}}, \bibnamefont{and}
  \bibinfo{author}{\bibfnamefont{V.~S.} \bibnamefont{Popov}},
  \bibinfo{journal}{Phys. Rev. Lett.} \textbf{\bibinfo{volume}{104}},
  \bibinfo{pages}{220404} (\bibinfo{year}{2010}).

\bibitem[{\citenamefont{Gelfer et~al.}(2015)\citenamefont{Gelfer, Mironov,
  Fedotov, Bashmakov, Nerush, Kostyukov, and Narozhny}}]{gelfer15}
\bibinfo{author}{\bibfnamefont{E.}~\bibnamefont{Gelfer}},
  \bibinfo{author}{\bibfnamefont{A.}~\bibnamefont{Mironov}},
  \bibinfo{author}{\bibfnamefont{A.}~\bibnamefont{Fedotov}},
  \bibinfo{author}{\bibfnamefont{V.}~\bibnamefont{Bashmakov}},
  \bibinfo{author}{\bibfnamefont{E.}~\bibnamefont{Nerush}},
  \bibinfo{author}{\bibfnamefont{I.~Y.} \bibnamefont{Kostyukov}},
  \bibnamefont{and} \bibinfo{author}{\bibfnamefont{N.}~\bibnamefont{Narozhny}},
  \bibinfo{journal}{Physical Review A} \textbf{\bibinfo{volume}{92}},
  \bibinfo{pages}{022113} (\bibinfo{year}{2015}).

\bibitem[{\citenamefont{Bell and Kirk}(2008)}]{kirk08}
\bibinfo{author}{\bibfnamefont{A.~R.} \bibnamefont{Bell}} \bibnamefont{and}
  \bibinfo{author}{\bibfnamefont{J.~G.} \bibnamefont{Kirk}},
  \bibinfo{journal}{Phys. Rev. Lett.} \textbf{\bibinfo{volume}{101}},
  \bibinfo{pages}{200403} (\bibinfo{year}{2008}).

\bibitem[{\citenamefont{Nerush et~al.}(2011)}]{nerush11}
\bibinfo{author}{\bibfnamefont{E.~N.} \bibnamefont{Nerush}}
  \bibnamefont{et~al.}, \bibinfo{journal}{Phys. Rev. Lett.}
  \textbf{\bibinfo{volume}{106}}, \bibinfo{pages}{035001}
  (\bibinfo{year}{2011}).

\bibitem[{\citenamefont{Elkina et~al.}(2011)}]{elkina11}
\bibinfo{author}{\bibfnamefont{N.~V.} \bibnamefont{Elkina}}
  \bibnamefont{et~al.}, \bibinfo{journal}{Phys. Rev. ST Accel. Beams}
  \textbf{\bibinfo{volume}{14}}, \bibinfo{pages}{054401}
  (\bibinfo{year}{2011}).

\bibitem[{\citenamefont{King et~al.}(2013)\citenamefont{King, Elkina, and
  Ruhl}}]{king13a}
\bibinfo{author}{\bibfnamefont{B.}~\bibnamefont{King}},
  \bibinfo{author}{\bibfnamefont{N.}~\bibnamefont{Elkina}}, \bibnamefont{and}
  \bibinfo{author}{\bibfnamefont{H.}~\bibnamefont{Ruhl}},
  \bibinfo{journal}{Phys. Rev. A} \textbf{\bibinfo{volume}{87}},
  \bibinfo{pages}{042117} (\bibinfo{year}{2013}).

\bibitem[{\citenamefont{Mironov et~al.}(2014)\citenamefont{Mironov, Narozhny,
  and Fedotov}}]{mironov14}
\bibinfo{author}{\bibfnamefont{A.~A.} \bibnamefont{Mironov}},
  \bibinfo{author}{\bibfnamefont{N.~B.} \bibnamefont{Narozhny}},
  \bibnamefont{and} \bibinfo{author}{\bibfnamefont{A.~M.}
  \bibnamefont{Fedotov}}, \bibinfo{journal}{Phys. Lett. A}
  \textbf{\bibinfo{volume}{378}}, \bibinfo{pages}{3254} (\bibinfo{year}{2014}).

\bibitem[{\citenamefont{King and Ruhl}(2013)}]{king13b}
\bibinfo{author}{\bibfnamefont{B.}~\bibnamefont{King}} \bibnamefont{and}
  \bibinfo{author}{\bibfnamefont{H.}~\bibnamefont{Ruhl}},
  \bibinfo{journal}{Phys. Rev. D} \textbf{\bibinfo{volume}{88}},
  \bibinfo{pages}{013005} (\bibinfo{year}{2013}).

\bibitem[{\citenamefont{Harvey et~al.}(2015)\citenamefont{Harvey, Ilderton, and
  King}}]{harvey15}
\bibinfo{author}{\bibfnamefont{C.~N.} \bibnamefont{Harvey}},
  \bibinfo{author}{\bibfnamefont{A.}~\bibnamefont{Ilderton}}, \bibnamefont{and}
  \bibinfo{author}{\bibfnamefont{B.}~\bibnamefont{King}},
  \bibinfo{journal}{Phys. Rev. A} \textbf{\bibinfo{volume}{91}},
  \bibinfo{pages}{013822} (\bibinfo{year}{2015}).

\bibitem[{\citenamefont{Kirk}(2016)}]{kirk16}
\bibinfo{author}{\bibfnamefont{J.~G.} \bibnamefont{Kirk}},
  \bibinfo{journal}{Plasma Physics and Controlled Fusion}
  \textbf{\bibinfo{volume}{58}}, \bibinfo{pages}{085005}
  (\bibinfo{year}{2016}).

\bibitem[{\citenamefont{Hu and Huang}(2015)}]{hu15}
\bibinfo{author}{\bibfnamefont{H.}~\bibnamefont{Hu}} \bibnamefont{and}
  \bibinfo{author}{\bibfnamefont{J.}~\bibnamefont{Huang}},
  \bibinfo{journal}{Phys. Rev.} \textbf{\bibinfo{volume}{A92}},
  \bibinfo{pages}{062105} (\bibinfo{year}{2015}).

\bibitem[{\citenamefont{Raicher et~al.}(2016)\citenamefont{Raicher, Eliezer,
  and Zigler}}]{raicher16}
\bibinfo{author}{\bibfnamefont{E.}~\bibnamefont{Raicher}},
  \bibinfo{author}{\bibfnamefont{S.}~\bibnamefont{Eliezer}}, \bibnamefont{and}
  \bibinfo{author}{\bibfnamefont{A.}~\bibnamefont{Zigler}},
  \bibinfo{journal}{arXiv preprint arXiv:1606.00476}  (\bibinfo{year}{2016}).

\bibitem[{\citenamefont{{Di Piazza}}(2016)}]{dipiazza16}
\bibinfo{author}{\bibfnamefont{A.}~\bibnamefont{{Di Piazza}}}
  (\bibinfo{year}{2016}), \eprint{arXiv:1608.08120}.

\bibitem[{\citenamefont{NIST}(2015)}]{nist_dlmf}
\bibinfo{author}{\bibnamefont{NIST}}, \emph{\bibinfo{title}{Nist digital
  library of mathematical functions}},
  \bibinfo{howpublished}{http://dlmf.nist.gov/} (\bibinfo{year}{2015}).

\bibitem[{\citenamefont{Elkina et~al.}(2014)\citenamefont{Elkina, Fedotov,
  Herzing, and Ruhl}}]{elkina14}
\bibinfo{author}{\bibfnamefont{N.~V.} \bibnamefont{Elkina}},
  \bibinfo{author}{\bibfnamefont{A.~M.} \bibnamefont{Fedotov}},
  \bibinfo{author}{\bibfnamefont{C.}~\bibnamefont{Herzing}}, \bibnamefont{and}
  \bibinfo{author}{\bibfnamefont{H.}~\bibnamefont{Ruhl}},
  \bibinfo{journal}{Phys. Rev. E} \textbf{\bibinfo{volume}{89}},
  \bibinfo{pages}{053315} (\bibinfo{year}{2014}).

\bibitem[{\citenamefont{Heinzl and Ilderton}(2009)}]{ilderton09}
\bibinfo{author}{\bibfnamefont{T.}~\bibnamefont{Heinzl}} \bibnamefont{and}
  \bibinfo{author}{\bibfnamefont{A.}~\bibnamefont{Ilderton}},
  \bibinfo{journal}{Opt. Commun.} \textbf{\bibinfo{volume}{282}},
  \bibinfo{pages}{1879} (\bibinfo{year}{2009}).

\bibitem[{Note1()}]{Note1}
Note1, \bibinfo{note}{in this paper, a different convention for the Volkov
  ansatz $\protect \qopname \relax o{exp}(+i p\cdot x)$ is used.}

\bibitem[{\citenamefont{M{\"u}ller-Kirsten}(2006)}]{muellerkirsten06}
\bibinfo{author}{\bibfnamefont{K.~J.~W.} \bibnamefont{M{\"u}ller-Kirsten}},
  \emph{\bibinfo{title}{Introduction to Quantum Mechanics}}
  (\bibinfo{publisher}{World Scientific}, \bibinfo{year}{2006}).

\bibitem[{\citenamefont{Bender and Orszag}(1978)}]{bender78}
\bibinfo{author}{\bibfnamefont{C.~M.} \bibnamefont{Bender}} \bibnamefont{and}
  \bibinfo{author}{\bibfnamefont{S.~A.} \bibnamefont{Orszag}},
  \emph{\bibinfo{title}{Advanced Mathematical Methods for Scientists and
  Engineers}} (\bibinfo{publisher}{Springer}, \bibinfo{year}{1978}).

\bibitem[{\citenamefont{Jeffreys}(1925)}]{jeffreys25}
\bibinfo{author}{\bibfnamefont{H.}~\bibnamefont{Jeffreys}},
  \bibinfo{journal}{Proc. London Math. Soc.}
  \textbf{\bibinfo{volume}{s2-23(1)}}, \bibinfo{pages}{428}
  (\bibinfo{year}{1925}).

\bibitem[{\citenamefont{Di~Piazza}(2014)}]{dipiazza14}
\bibinfo{author}{\bibfnamefont{A.}~\bibnamefont{Di~Piazza}},
  \bibinfo{journal}{Phys. Rev. Lett.} \textbf{\bibinfo{volume}{113}},
  \bibinfo{pages}{040402} (\bibinfo{year}{2014}).

\bibitem[{\citenamefont{Di~Piazza}(2015)}]{dipiazza15}
\bibinfo{author}{\bibfnamefont{A.}~\bibnamefont{Di~Piazza}},
  \bibinfo{journal}{Phys. Rev. A} \textbf{\bibinfo{volume}{91}},
  \bibinfo{pages}{042118} (\bibinfo{year}{2015}).

\bibitem[{\citenamefont{Itzykson and Zuber}(1980)}]{itzykson80}
\bibinfo{author}{\bibfnamefont{C.}~\bibnamefont{Itzykson}} \bibnamefont{and}
  \bibinfo{author}{\bibfnamefont{J.-B.} \bibnamefont{Zuber}},
  \emph{\bibinfo{title}{Quantum Field Theory}}
  (\bibinfo{publisher}{McGraw-Hill}, \bibinfo{address}{New York},
  \bibinfo{year}{1980}).

\bibitem[{\citenamefont{Harvey et~al.}(2012)\citenamefont{Harvey, Heinzl,
  Ilderton, and Marklund}}]{harvey12}
\bibinfo{author}{\bibfnamefont{C.}~\bibnamefont{Harvey}},
  \bibinfo{author}{\bibfnamefont{T.}~\bibnamefont{Heinzl}},
  \bibinfo{author}{\bibfnamefont{A.}~\bibnamefont{Ilderton}}, \bibnamefont{and}
  \bibinfo{author}{\bibfnamefont{M.}~\bibnamefont{Marklund}},
  \bibinfo{journal}{Phys. Rev. Lett.} \textbf{\bibinfo{volume}{109}},
  \bibinfo{pages}{100402} (\bibinfo{year}{2012}).

\bibitem[{\citenamefont{Jackson}(1975)}]{jackson75}
\bibinfo{author}{\bibfnamefont{J.~D.} \bibnamefont{Jackson}},
  \emph{\bibinfo{title}{Classical Electrodynamics}} (\bibinfo{publisher}{John
  Wiley \& Sons, Inc.}, \bibinfo{address}{New York}, \bibinfo{year}{1975}).

\bibitem[{Note2()}]{Note2}
Note2, \bibinfo{note}{in the usual discussion of the validity of the plane wave
  model $\eta $ is replaced with $\chi = \xi \eta $, but the reasoning is the
  same.}

\bibitem[{\citenamefont{Berestetskii et~al.}(1982)\citenamefont{Berestetskii,
  Lifshitz, and Pitaevskii}}]{landau4}
\bibinfo{author}{\bibfnamefont{V.~B.} \bibnamefont{Berestetskii}},
  \bibinfo{author}{\bibfnamefont{E.~M.} \bibnamefont{Lifshitz}},
  \bibnamefont{and} \bibinfo{author}{\bibfnamefont{L.~P.}
  \bibnamefont{Pitaevskii}}, \emph{\bibinfo{title}{Quantum Electrodynamics
  (second edition)}} (\bibinfo{publisher}{Butterworth-Heinemann},
  \bibinfo{address}{Oxford}, \bibinfo{year}{1982}).

\bibitem[{\citenamefont{Bauer et~al.}(1995)\citenamefont{Bauer, Mulser, and
  Steeb}}]{bauer95}
\bibinfo{author}{\bibfnamefont{D.}~\bibnamefont{Bauer}},
  \bibinfo{author}{\bibfnamefont{P.}~\bibnamefont{Mulser}}, \bibnamefont{and}
  \bibinfo{author}{\bibfnamefont{W.~H.} \bibnamefont{Steeb}},
  \bibinfo{journal}{Phys. Rev. Lett.} \textbf{\bibinfo{volume}{75}},
  \bibinfo{pages}{4622} (\bibinfo{year}{1995}).

\bibitem[{\citenamefont{Kaplan and Pokrovsky}(2005)}]{kaplan05}
\bibinfo{author}{\bibfnamefont{A.~E.} \bibnamefont{Kaplan}} \bibnamefont{and}
  \bibinfo{author}{\bibfnamefont{A.~L.} \bibnamefont{Pokrovsky}},
  \bibinfo{journal}{Phys. Rev. Lett.} \textbf{\bibinfo{volume}{95}},
  \bibinfo{pages}{053601} (\bibinfo{year}{2005}).

\bibitem[{\citenamefont{Gonoskov et~al.}(2014)\citenamefont{Gonoskov, Bashinov,
  Gonoskov, Harvey, Ilderton, Kim, Marklund, Mourou, and Sergeev}}]{gonoskov14}
\bibinfo{author}{\bibfnamefont{A.}~\bibnamefont{Gonoskov}},
  \bibinfo{author}{\bibfnamefont{A.}~\bibnamefont{Bashinov}},
  \bibinfo{author}{\bibfnamefont{I.}~\bibnamefont{Gonoskov}},
  \bibinfo{author}{\bibfnamefont{C.}~\bibnamefont{Harvey}},
  \bibinfo{author}{\bibfnamefont{A.}~\bibnamefont{Ilderton}},
  \bibinfo{author}{\bibfnamefont{A.}~\bibnamefont{Kim}},
  \bibinfo{author}{\bibfnamefont{M.}~\bibnamefont{Marklund}},
  \bibinfo{author}{\bibfnamefont{G.}~\bibnamefont{Mourou}}, \bibnamefont{and}
  \bibinfo{author}{\bibfnamefont{A.}~\bibnamefont{Sergeev}},
  \bibinfo{journal}{Phys. Rev. Lett.} \textbf{\bibinfo{volume}{113}},
  \bibinfo{pages}{014801} (\bibinfo{year}{2014}),
  \urlprefix\url{http://link.aps.org/doi/10.1103/PhysRevLett.113.014801}.

\end{thebibliography}

\end{document}